\documentclass[11pt]{article}
\usepackage{setspace}
\usepackage{adjustbox}
\usepackage{color}
\usepackage{amsmath, amsthm, amssymb}
\usepackage[left=1in,top=1in,right=1in,bottom=1in,nohead]{geometry}
\usepackage{multirow}
\usepackage{cite}
\theoremstyle{definition}

\theoremstyle{definition}

\theoremstyle{definition}

\title{A model-based approach for identifying signatures of balancing selection in genetic data}
\author{Michael DeGiorgio$^1$, Kirk E. Lohmueller$^{1,*}$, and Rasmus Nielsen$^{1,2,3}$}
\date{}
\begin{document}
    \doublespacing
    \maketitle
    \begin{flushleft}
        $^1$Department of Integrative Biology, University of California, Berkeley, CA, USA\\
        $^2$Department of Statistics, University of California, Berkeley, CA, USA\\
        $^3$Department of Biology, University of Copenhagen, Copenhagen, Denmark\\[2ex]
        $^*$\textsl{Present address}: Department of Ecology and Evolutionary Biology, University of California, Los Angeles, CA, USA\\[10ex]
    \end{flushleft}


    \begin{flushleft}
        \textbf{Corresponding author}\\
        Michael DeGiorgio\\
        Department of Integrative Biology\\
        University of California\\
        $4134$ Valley Life Sciences Building\\
        Berkeley, CA 94720, USA\\
        Phone +1 510 643 0060\\
        email: mdegiorgio@berkeley.edu\\[2ex]
        \textbf{Running Head:} Detecting balancing selection using the coalescent\\[2ex]
        \textbf{Classification:} Biological Sciences, Genetics
    \end{flushleft}

    \clearpage

    \begin{abstract}
        While much effort has focused on detecting positive and negative directional selection in the human genome, relatively little work has been devoted to balancing selection. This lack of attention is likely due to the paucity of sophisticated methods for identifying sites under balancing selection. Here we develop two composite likelihood ratio tests for detecting balancing selection. Using simulations, we show that these methods outperform competing methods under a variety of assumptions and demographic models. We apply the new methods to whole-genome human data, and find a number of previously-identified loci with strong evidence of balancing selection, including several HLA genes. Additionally, we find evidence for many novel candidates, the strongest of which is \textsl{FANK1}, an imprinted gene that suppresses apoptosis, is expressed during meiosis in males, and displays marginal signs of segregation distortion. We hypothesize that balancing selection acts on this locus to stabilize the segregation distortion and negative fitness effects of the distorter allele. Thus, our methods are able to reproduce many previously-hypothesized signals of balancing selection, as well as discover novel interesting candidates.
    \end{abstract}

    \clearpage

    \begin{sloppypar}
        \section*{Introduction}
            Balancing selection maintains variation within a population. Multiple processes can lead to balancing selection. In overdominance, the heterozygous genotype has higher fitness than either of the homozygous genotypes \cite{Fisher22, Andres11}. In frequency-dependent balancing selection, the fitness of an allele is inversely related to its frequency in the population \cite{WilsonAndTurelli86, Andres11}. In a fluctuating or spatially-structured environment, balancing selection can occur when different alleles are favored in different environments over time or geography \cite{Levene53, Nagylaki75, Andres11}. Finally, balancing selection can also be a product of opposite directed effects of segregation distortion balanced by negative selection against the distorter \cite{CharlesworthAndCharlesworth10}. That is, segregation distortion leads to one allele increasing in frequency. However, if that allele is deleterious, then it is reduced in frequency by negative selection. The combined effect of these opposing forces leads to a blanaced polymorphism.

            The genetic signatures of long-term balancing selection at a locus can roughly be divided into three categories \cite{Andres11}. The first signature is that the distribution of allele frequencies will be enriched for intermediate frequency alleles. This occurs because the selected locus itself is likely at moderate frequency within the population and, thus, neutral linked loci will also be at intermediate frequency. The second signature is the presence of trans-specific polymorphisms, which are polymorphisms that are shared among species \cite{SegurelEtAl12}. This is a result of alleles being maintained over long evolutionary time periods, sometimes for millions of years \cite{KleinEtAl93, KleinEtAl98, KleinEtAl07}. The third signature is an increased density of polymorphic sites. This is due to neutral loci sharing similar deep genealogies as that of the linked selected site, increasing the probability of observing mutations at the neutral loci.

            \textcolor{black}{The majority of selection scans in humans have focused on positive and negative directional selection. These studies have found evidence of both types of selection, with negative selection being ubiquitous, and the amount and mechanism of positive selection currently being debated \cite{HernandezEtAl11, LohmuellerEtAl11PLoS, GrankaEtAl12}. However, it is unclear how much balancing selection exists in the human genome. Some scans for balancing selection (e.g., Bubb~\textsl{et al.} \cite{BubbEtAl06} and Andr\'es~\textsl{et al.} \cite{AndresEtAl09}) have been carried out using summary statistics such as the Hudson-Kreitman-Aguad\'e (HKA) test \cite{HudsonEtAl87} and Tajima's $D$ \cite{Tajima89} as well as combinations of summary statistics \cite{Innan06, AndresEtAl09} (though see S\'{e}gural~\textsl{et al}. \cite{SegurelEtAl12} and Leffler~\textsl{et al.} \cite{LefflerEtAl13} for recent complementary approaches). The power of such approaches in unclear, and so it is uncertain how important balancing selection is in the human genome. Because balancing selection shapes the genealogy of a sample around a selected locus, more power can be gained by implementing a model of the genealogical process under balancing selection \cite{KaplanEtAl88, HudsonAndKaplan88}. Composite likelihood methods have proven to be extremely useful for the analysis of genetic variation data using complex population genetic models. \cite{Hudson01, KimAndStephan02, KimAndNielsen04, JensenEtAl05, NielsenEtAl05GenomeRes, NielsenEtAl09, ChenEtAl10}. This approach allows estimation under models without requiring full likelihood calculations, permitting many complex models to be investigated.}


            In this article, we develop two composite likelihood ratio methods to detect balancing selection, which we denote by $T_1$ and $T_2$. These methods are based on modeling the effect of balancing selection on the genealogy at linked neutral loci (e.g., Kaplan~\textsl{et al.}~(1988)\cite{KaplanEtAl88} and Hudson~and~Kaplan~(1988)\cite{HudsonAndKaplan88}) and take into consideration the spatial distributions of polymorphisms and substitutions around a selected site. Through simulations, we show that our methods outperform both HKA and Tajima's $D$ under a variety of demographic assumptions. Further, we apply our methods to autosomal whole-genome sequencing data consisting of nine unrelated European (CEU) and nine unrelated African (YRI) individuals. We find support for multiple targets of balancing selection in the human genome, including previously hypothesized regions such as the human leukocyte antigen (HLA) locus. Additionally, we find evidence for balancing selection at the \textsl{FANK1} gene, which we hypothesize to result from segregation distortion.

        \section*{Results}
            \subsection*{Theory}
                \subsubsection*{A new test for balancing selection}
                    \textcolor{black}{In this section, we provide a basic overview of a new test for balancing selection, and we describe the method in greater detail in the \textsl{Kaplan-Darden-Hudson model}, \textsl{Solving the recursion relation}, \textsl{A composite likelihood ratio test based on polymorphism and substitution}, and \textsl{A composite likelihood ratio test based on frequency spectra and substitutions} sections.} We have developed a new statistical method for detecting balancing selection, which is based on the model of Kaplan, Darden, and Hudson \cite{KaplanEtAl88, HudsonAndKaplan88} (full details provided in the \textsl{Kaplan-Darden-Hudson model} section). Under this model, we calculate the expected distribution of allele frequencies using simulations, and approximate the probability of observing a fixed difference or polymorphism at a site as a function of its genomic distance to a putative site under balancing selection. Using these calculations, we construct composite likelihood tests that can be used to identify sites under balancing selection, similar to the approaches by Kim~and~Stephan \cite{KimAndStephan02} and Nielsen~\textsl{et al.} \cite{NielsenEtAl05GenomeRes} for detecting selective sweeps.

                \subsubsection*{Basic framework}
\begin{figure*}[h]
            \centering
            \includegraphics[width=1.0\textwidth]{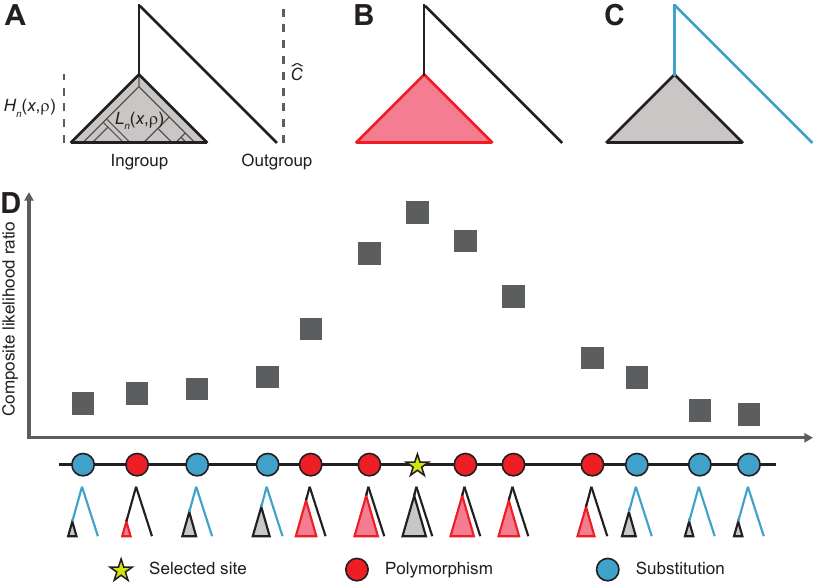}
            \caption{Calculation of probabilities of polymorphism and substitution under balancing selection and the incorporation of these probabilities into a genome scan. (\textsl{A}) Relationship among tree length $L_n(x, \rho)$, tree height $H_n(x, \rho)$ and inter-specific coalescence time $\widehat{C}$. (\textsl{B}) A site is polymorphic if a mutation occurred on the $L_n(x, \rho)$ length of branches until the most recent common ancestor of the ingroup sample (red region). (\textsl{C}) A site is a substitution if a mutation occurred on the $2\widehat{C} - H_n(x, \rho)$ length of branches that represent the divergence between the outgroup species and the most recent common ancestor of the ingroup species (blue region). (\textsl{D}) Height and length of genealogies in relationship to their spatial proximity to a selected site and how the shapes of these genealogies affect the pattern of polymorphism around the site. The composite likelihood ratio is high near a selected site as there is an excess of polymorphisms close to the site and a deficit far from the site.}
            \label{figure:method_illustration}
        \end{figure*}

                    Consider a biallelic site $S$ that is under strong balancing selection and maintains an allele $A_1$ at frequency $x$ and an allele $A_2$ at frequency $1 - x$. Consider a neutral locus $i$ that is linked to the selected locus $S$. Denote the scaled recombination rate between the selected locus and the neutral locus as $\rho_i = 2Nr_i$, where $N$ is the diploid population size and $r_i$ is the per-generation recombination rate. Assume we have a sample of $n$ genomes from an ingroup species (e.g., humans) and a single genome from an outgroup species (e.g., chimpanzee). From these data, we can estimate the genome-wide expected coalescence time $\widehat{C}$ between the ingroup and outgroup species (see \textsl{Materials and Methods} for details). Also, under the Kaplan-Darden-Hudson model, we can obtain the expected tree length $L_n(x, \rho)$ and height $H_n(x, \rho)$ for a sample of $n$ lineages affected by balancing selection by solving a set of recursive equations using the numerical approach described in the \textsl{Solving the recursion relation}. The relationship among $\widehat{C}$, $L_n(x, \rho)$, and $H_n(x, \rho)$ is depicted in Figure~\ref{figure:method_illustration}\textsl{A}. Assuming a small mutation rate, the probability that a site is polymorphic under a model of balancing selection, given that it contains either a polymorphism or a substitution (fixed difference), is
                    \begin{equation}\label{eq:probPolymorphism}
                        p_{n, \rho, x} = \frac{L_n(x, \rho)}{2\widehat{C} - H_n(x, \rho) + L_n(x, \rho)},
                    \end{equation}
                    and the conditional probability that it contains a substitution is $s_{n, \rho, x} = 1 - p_{n, \rho, x}$. That is, conditional on a mutation occurring on the genealogy relating the $n$ ingroup genomes and the outgroup genome, the probability that a site is polymorphic is the probability that a mutation occurs before the most recent common ancestor of the $n$ ingroup species (\textsl{i.e.,} mutation occurs on red branches indicated in Fig.~\ref{figure:method_illustration}\textsl{B}), and the probability that a site contains a substitution is the probability that a mutation occurs along the branch leading from the outgroup sequence to the most recent common ancestor of the $n$ ingroup species (\textsl{i.e.,} mutation occurs on blue branches indicated in Fig.~\ref{figure:method_illustration}\textsl{C}).

                    Figure~\ref{figure:method_illustration}\textsl{D} shows how the spatial distribution of polymorphism around a selected site is influenced by the underlying genealogy at the site and how this spatial distribution of polymorphism can be used to provide evidence for balancing selection. Within a window of sites, we can obtain the composite likelihood that a particular site is under selection by multiplying the conditional probability of observing a polymorphism or a substitution at every other neutral site as a function of the distance of the neutral site to the balanced polymorphism.

                \subsubsection*{Kaplan-Darden-Hudson model}
                    The genealogy of a neutral locus $i$ linked to the selected locus $S$ can be traced back in time using the Kaplan, Darden, and Hudson \cite{KaplanEtAl88, HudsonAndKaplan88} model, which provides a framework for modeling the coalescent process at a neutral locus that is linked to a locus under balancing selection. Their framework involves modeling selection as a structured population containing two demes representing each of the two allelic classes and migration taking the role of recombination and mutation. Lineages within the first deme are linked to $A_1$ alleles and lineages within the second deme are linked to $A_2$ alleles. Lineages migrate between demes by changing their genomic background. That is, a lineage in the first deme will migrate to the second deme if there was a mutation that changed an $A_1$ allele to an $A_2$ allele or if there was a recombination event that transferred a lineage linked to an $A_1$ allele to an $A_2$ background. Similarly, a lineage in the second deme will migrate to the first deme if there was a mutation that changed an $A_2$ allele to an $A_1$ allele or if there was a recombination event that transferred a lineage linked to an $A_2$ allele to an $A_1$ background. The rate at which a lineage linked to an $A_1$ background transfers to an $A_2$ background is $\beta_1 = \theta_1 + \rho_i (1 - x)$ and the rate at which a lineage linked to an $A_2$ background transfers to an $A_1$ background is $\beta_2 = \theta_2 + \rho_i x$.

                    Consider a sample of $n$ lineages with $k$ lineages linked to allele $A_1$ (\textsl{i.e.}, in the first deme) and $n - k$ lineages linked to allele $A_2$ (\textsl{i.e.}, in the second deme). Given this configuration, only four events are possible. The first event involves a coalescence of a pair of lineages linked to $A_1$ alleles, the second involves a coalescence of a pair of lineages linked to $A_2$ alleles, the third involves the transfer of a lineage from an $A_1$ background to an $A_2$ background, and the fourth involves the transfer of a lineage from an $A_2$ background to an $A_1$ background. The time until the first event (\textsl{i.e.}, a coalescence or a transfer of background) is exponentially distributed with rate
                    \begin{equation}\label{eq:rate_next_event}
                        \lambda_{k, n - k}(x, \rho) = \frac{\binom{k}{2}}{x} + \frac{\binom{n-k}{2}}{1-x} + \frac{k\beta_2(1-x)}{x} + \frac{(n-k)\beta_1x}{1-x}.
                    \end{equation}
                    The probability that the event is a coalescence of a pair of $A_1$-linked lineages is
                    \begin{equation}\label{eq:prob_coal_1}
                        c_{k, n - k}^{(1)}(x, \rho) = \frac{\binom{k}{2}}{x\lambda_{k,n-k}(x, \rho)},
                    \end{equation}
                    the event is a coalescence of a pair of $A_2$-linked lineages is
                    \begin{equation}\label{eq:prob_coal_2}
                        c_{k, n - k}^{(2)}(x, \rho) = \frac{\binom{n-k}{2}}{(1-x)\lambda_{k,n-k}(x, \rho)},
                    \end{equation}
                    the event is a transfer from an $A_1$ to an $A_2$ background is
                    \begin{equation}\label{eq:prob_mig_1}
                        m_{k, n - k}^{(1)}(x, \rho) = \frac{k\beta_2(1-x)}{x\lambda_{k,n-k}(x, \rho)},
                    \end{equation}
                    and the event is a transfer from an $A_2$ to an $A_1$ background is
                    \begin{equation}\label{eq:prob_mig_2}
                        m_{k, n - k}^{(2)}(x, \rho) = \frac{(n-k)\beta_1x}{(1-x)\lambda_{k,n-k}(x, \rho)}.
                    \end{equation}
                    Note that in the notation of Kaplan~\textsl{et al.}~(1988) \cite{KaplanEtAl88}, $\lambda_{k,n-k}(x, \rho) = h_{k,n-k}(x)$, $c_{k,n-k}^{(1)}(x, \rho) = q_{k-1,n-k}(x)$, $c_{k,n-k}^{(2)}(x, \rho) = q_{k,n-k-1}(x)$, $m_{k,n-k}^{(1)}(x, \rho) = q_{k-1,n-k+1}(x)$, and $m_{k,n-k}^{(2)}(x, \rho) = q_{k+1,n-k-1}(x)$.

                    Let $L_{k,n-k}(x, \rho)$ denote the expected tree length given a sample with $k$ $A_1$-linked lineages and $n - k$ $A_2$-linked lineages. Using eq.~18 of Kaplan~\textsl{et al.}~(1988) \cite{KaplanEtAl88}, the expected total tree length can be expressed using the recursion relation
                    \begin{align}\label{eq:treeLengthRecursion}
                        L_{k,n-k}(x, \rho) &= \frac{n}{\lambda_{k,n-k}(x, \rho)} + c_{k,n-k}^{(1)}(x, \rho)L_{k-1,n-k}(x, \rho) + c_{k,n-k}^{(2)}(x, \rho)L_{k,n-k-1}(x, \rho)\notag\\
                        &\phantom{xx}+ m_{k,n-k}^{(1)}(x, \rho)L_{k-1,n-k+1}(x, \rho) + m_{k,n-k}^{(2)}(x, \rho)L_{k+1,n-k-1}(x, \rho).
                    \end{align}
                    Similarly, the expected tree height $H_{k,n-k}(x, \rho)$ given a sample with $k$ $A_1$-linked lineages and $n - k$ $A_2$-linked lineages can be expressed by
                    \begin{align}\label{eq:treeHeightRecursion}
                        H_{k,n-k}(x, \rho) &= \frac{1}{\lambda_{k,n-k}(x, \rho)} + c_{k,n-k}^{(1)}(x, \rho)H_{k-1,n-k}(x, \rho) + c_{k,n-k}^{(2)}(x, \rho)H_{k,n-k-1}(x, \rho)\notag\\
                        &\phantom{xx}+ m_{k,n-k}^{(1)}(x, \rho)H_{k-1,n-k+1}(x, \rho) + m_{k,n-k}^{(2)}(x, \rho)H_{k+1,n-k-1}(x, \rho).
                    \end{align}

                \subsubsection*{Solving the recursion relation}
                    Consider a sample of $n$ lineages. Denote the $(n + 1)$-dimensional vector of tree lengths for a sample of size $n$ as
                    \[ \boldsymbol{\ell}^{(n)} =
                        \begin{bmatrix}
                            L_{0, n}(x, \rho)\\[0.5em]
                            L_{1, n-1}(x, \rho)\\[0.5em]
                            L_{2, n-2}(x, \rho)\\[0.5em]
                            \vdots\\[0.5em]
                            L_{n, 0}(x, \rho)
                        \end{bmatrix},
                    \]
                    such that element $k$, $k = 0, 1, \ldots, n$, of $\boldsymbol{\ell}^{(n)}$ is $\boldsymbol{\ell}_k^{(n)} = L_{k, n - k}(x, \rho)$. Next, define the ($n + 1$)-dimensional vector
                    \[ \mathbf{b}^{(n)} =
                        \begin{bmatrix}
                            \frac{n}{\lambda_{0,n}(x, \rho)} + c_{0,n}^{(2)}(x, \rho)L_{0, n-1}(x, \rho)\\[0.5em]
                            \frac{n}{\lambda_{1,n-1}(x, \rho)} + c_{1,n-1}^{(1)}(x, \rho)L_{0, n-1}(x, \rho) + c_{1,n-1}^{(2)}(x, \rho)L_{1, n-2}(x, \rho)\\[0.5em]
                            \frac{n}{\lambda_{2,n-2}(x, \rho)} + c_{2,n-2}^{(1)}(x, \rho)L_{1, n-2}(x, \rho) + c_{2,n-2}^{(2)}(x, \rho)L_{2, n-3}(x, \rho)\\[0.5em]
                            \vdots\\[0.5em]
                            \frac{n}{\lambda_{n,0}(x, \rho)} + c_{n,0}^{(1)}(x, \rho)L_{n-1, 0}(x, \rho)
                        \end{bmatrix},
                    \]
                    such that element 0 is
                    \begin{displaymath}
                        \mathbf{b}_0^{(n)} = \frac{n}{\lambda_{0,n}(x, \rho)} + c_{0,n}^{(2)}(x, \rho)\boldsymbol{\ell}_{0}^{(n - 1)},
                    \end{displaymath}
                    element $n$ is
                    \begin{displaymath}
                        \mathbf{b}_n^{(n)} = \frac{n}{\lambda_{n,0}(x, \rho)} + c_{n,0}^{(1)}(x, \rho)\boldsymbol{\ell}_{n - 1}^{(n - 1)},
                    \end{displaymath}
                    and element $k$, $k = 1, 2, \ldots, n-1$ is
                    \begin{displaymath}
                        \mathbf{b}_k^{(n)} = \frac{n}{\lambda_{k,n-k}(x, \rho)} + c_{k,n - k}^{(1)}(x, \rho)\boldsymbol{\ell}_{k - 1}^{(n - 1)} + c_{k,n - k}^{(2)}(x, \rho)\boldsymbol{\ell}_{k}^{(n - 1)}.
                    \end{displaymath}
                    Further, consider an $(n + 1)\times(n + 1)$-dimensional tridiagonal matrix of migration rates
                    \[ \mathbf{M}^{(n)} =
                        \begin{bmatrix}
                            1 & -m_{0,n}^{(2)}(x, \rho) & 0 & 0 & 0\\[0.5em]
                            -m_{1,n-1}^{(1)}(x, \rho) & 1 & -m_{1,n-1}^{(2)}(x, \rho) & 0 & 0\\[0.5em]
                            0 & -m_{2,n-2}^{(1)}(x, \rho) & 1 & \ddots & 0\\[0.5em]
                            0 & 0 & \ddots & \ddots & -m_{n-1,1}^{(2)}(x, \rho)\\[0.5em]
                            0 & 0 & 0 & -m_{n,0}^{(1)}(x, \rho) & 1
                        \end{bmatrix},
                    \]
                    with $(n + 1)$-dimensional main diagonal $\text{diag}(\mathbf{M}^{(n)}) = [1, 1, \ldots, 1]$, $n$-dimensional lower diagonal $\text{lower}(\mathbf{M}^{(n)}) = [-m_{1,n-1}^{(1)}(x, \rho), -m_{2,n-2}^{(1)}(x, \rho), \ldots, -m_{n,0}^{(1)}(x, \rho)]$, and $n$-dimensional upper diagonal $\text{upper}(\mathbf{M}^{(n)}) = [-m_{0,n}^{(2)}(x, \rho), -m_{1,n-1}^{(2)}(x, \rho), \ldots, -m_{n-1,1}^{(2)}(x, \rho)]$. All elements that do not fall on the main, lower, and upper diagonals of $\mathbf{M}^{(n)}$ are zero.

                    Given $\mathbf{M}^{(n)}$, $\mathbf{b}^{(n)}$, and $\boldsymbol{\ell}^{(n)}$, we can rewrite the recursion relation in eq.~\ref{eq:treeLengthRecursion} as system of equations
                    \begin{equation}\label{eq:treeLengthMatrixForm}
                        \mathbf{M}^{(n)}\boldsymbol{\ell}^{(n)} = \mathbf{b}^{(n)}.
                    \end{equation}
                    Because we can calculate eqs.~\ref{eq:prob_mig_1}~and~\ref{eq:prob_mig_2}, $\mathbf{M}^{(n)}$ is a constant matrix. For a sample of size $n$, suppose we know $\boldsymbol{\ell}^{(n-1)}$ for a sample of size $n - 1$. Therefore, $\boldsymbol{\ell}^{(n-1)}$ is now a constant vector and hence, because we can calculate eqs.~\ref{eq:rate_next_event}-\ref{eq:prob_coal_2}, $\mathbf{b}^{(n)}$ is also a constant vector. Therefore, eq.~\ref{eq:treeLengthMatrixForm} is a tridiagonal system of $n + 1$ equations with $n + 1$ unknowns, which can be solved in $O(n)$ time using the tridiagonal matrix algorithm \cite{Thomas49}.

                    The base case for the recursion in eq.~\ref{eq:treeHeightRecursion} is when the number of lineages equals one. That is, when all lineages have coalesced and the most recent common ancestor is linked either to an $A_1$ allele or to an $A_2$ allele. This base case can be represented by $L_{0,1}(x, \rho) = 0$ and $L_{1,0}(x, \rho) = 0$. Given these values, set $\boldsymbol{\ell}^{(1)} = [L_{0,1}(x, \rho), L_{1,0}(x, \rho)] = [0, 0]$ and solve the system of equations $\mathbf{M}^{(2)}\boldsymbol{\ell}^{(2)} = \mathbf{b}^{(2)}$ for $\boldsymbol{\ell}^{(2)}$. Next, given $\boldsymbol{\ell}^{(2)}$, solve the system of equations $\mathbf{M}^{(3)}\boldsymbol{\ell}^{(3)} = \mathbf{b}^{(3)}$ for $\boldsymbol{\ell}^{(3)}$. Iterate this processes until $\mathbf{M}^{(n)}\boldsymbol{\ell}^{(n)} = \mathbf{b}^{(n)}$ is solved for $\boldsymbol{\ell}^{(n)}$. An analogous process can be used to solve the recursion (eq.~\ref{eq:treeHeightRecursion}) for the expected tree height.

                    Using the framework in this section for a sample of size $n$, we can obtain values for $L_{0,n}(x, \rho), L_{1,n-1}(x, \rho), \ldots, L_{n,0}(x, \rho)$. Given that the $A_1$ allele has frequency $x$ and the $A_2$ allele has frequency $1 - x$, the expected tree length for a sample of size $n$ is
                    \begin{equation}\label{eq:expectedTreeLength}
                        L_n(x,\rho) = \sum_{k = 0}^n \binom{n}{k} x^k (1 - x)^{n - k} L_{k, n- k}(x, \rho).
                    \end{equation}
                    Similarly, we can obtain the expected tree height $H_n(x, \rho)$ for a sample of size $n$. The tree heights and total branch lengths are then used in eq.~\ref{eq:probPolymorphism} to compute the likelihood of the data under the selection model.

                \subsubsection*{A composite likelihood ratio test based on polymorphism and substitution}
                    In this section, we illustrate how eq.~\ref{eq:probPolymorphism} can be incorporated into a composite likelihood. We will then describe a likelihood ratio test that compares the balancing selection model described above to a neutral model based on the background genome patterns of polymorphism. Consider a window of $I$ sites that are either polymorphisms or substitutions and consider a putatively selected site $S$ located within the window. Suppose site $i$ within the window has $n_i$ sampled alleles, $a_i$ observed ancestral alleles, and is a recombination distance of $\rho_i$ from $S$. Let $\mathbf{n} = [n_1, n_2, \ldots, n_I]$, $\mathbf{a} = [a_1, a_2, \ldots, a_I]$, and $\boldsymbol{\rho} = [\rho_1, \rho_2, \ldots, \rho_I]$. Define the indicator random variable $\mathbf{1}_{\{a_i = k\}}$ that site $i$ has $k$ ancestral alleles. Using the Kaplan-Darden-Hudson model, the probability that site $i$ is polymorphic is $p_{n_i, \rho_i, x}$ and the probability that the site is a substitution (or fixed difference) is $s_{n_i, \rho_i, x} = 1 - p_{n_i, \rho_i, x}$. Under the model, the composite likelihood that site $S$ is under balancing selection is
                    \begin{equation}
                        \mathcal{L}_{\text{M}}(\mathbf{n}, \boldsymbol{\rho}, x \,;\, \mathbf{a}) = \prod_{i = 1}^I\Bigg[s_{n_i, \rho_i, x}\mathbf{1}_{\{a_i = 0\}} + p_{n_i,\rho_i, x}\sum_{k = 1}^{n_i - 1}\mathbf{1}_{\{a_i = k\}}\Bigg],
                    \end{equation}
                    which is maximized at $\widehat{x} = {{\arg\max} \atop {x \in (0,1)}} \mathcal{L}_{\text{M}}(\mathbf{n}, \boldsymbol{\rho}, x \,;\, \mathbf{a})$. Further, suppose that for a sample of size $k$, $k = 2, 3, \ldots, n$, conditioning only on sites that are polymorphisms or substitutions, the proportion of loci across the genome that are polymorphic is $\widehat{p}_k$ and the proportion of loci that are substitutions is $\widehat{s}_k = 1 - \widehat{p}_k$. Then the composite likelihood that site $S$ is evolving neutrally is
                    \begin{equation}
                        \mathcal{L}_{\text{B}}(\mathbf{n} \,;\, \mathbf{a}) = \prod_{i = 1}^I\Bigg[\widehat{s}_{n_i}\mathbf{1}_{\{a_i = 0\}} + \widehat{p}_{n_i}\sum_{k = 1}^{n_i - 1}\mathbf{1}_{\{a_i = k\}}\Bigg].
                    \end{equation}
                    It follows that the composite likelihood ratio test statistic that site $S$ is under balancing selection is $T_1 = 2\{\ln[\mathcal{L}_{\text{M}}(\mathbf{n}, \boldsymbol{\rho}, \widehat{x} \,;\, \mathbf{a})] - \ln[\mathcal{L}_{\text{B}}(\mathbf{n} \,;\, \mathbf{a})]\}$.

                \subsubsection*{A composite likelihood ratio test based on frequency spectra and substitutions}
                    A balanced polymorphism not only increases the number of polymorphisms at linked neutral sites, but also leads to an increase in allele frequencies at these sites. Therefore, power can be gained by using frequency spectra information in addition to information on the density of polymorphisms and substitutions.

                    Given a sample of size $n$, an $A_1$ allele at frequency $x$, $A_2$ allele at frequency $1 - x$, and a polymorphic neutral site that is $\rho$ recombination units from a selected site, we can obtain the probability $p_{n, k, \rho, x}$ that there are $k$, $k = 1, 2, \ldots, n - 1$, ancestral alleles observed at the neutral site. The composite likelihood that site $S$ is under balancing selection is
                    \begin{equation}
                        \mathcal{L}_{\text{M}}(\mathbf{n}, \boldsymbol{\rho}, x \,;\, \mathbf{a}) = \prod_{i = 1}^I\Bigg[s_{n_i,\rho_i, x}\mathbf{1}_{\{a_i = 0\}} + p_{n_i, \rho_i, x}\sum_{k = 1}^{n_i - 1}p_{n_i,k,\rho_i, x}\mathbf{1}_{\{a_i = k\}}\Bigg],
                    \end{equation}
                    which is maximized at $\widehat{x} = {{\arg\max} \atop {x \in (0,1)}} \mathcal{L}_{\text{M}}(\mathbf{n}, \boldsymbol{\rho}, x \,;\, \mathbf{a})$.
                    Further, suppose that for a sample of size $k$, $k = 2, 3, \ldots, n$, conditioning only on sites that are polymorphisms or substitutions, the proportion of polymorphic loci across the genome that have $j$, $j = 1, 2, \ldots, k - 1$, ancestral alleles is $\widehat{p}_{k,j}$. Then the composite likelihood that site $S$ is evolving neutrally is
                    \begin{equation}
                        \mathcal{L}_{\text{B}}(\mathbf{n} \,;\, \mathbf{a}) = \prod_{i = 1}^I\Bigg[\widehat{s}_{n_i}\mathbf{1}_{\{a_i = 0\}} + \widehat{p}_{n_i}\sum_{k = 1}^{n_i - 1}\widehat{p}_{n_i, k}\mathbf{1}_{\{a_i = k\}}\Bigg].
                    \end{equation}
                    It follows that the composite likelihood ratio test statistic that site $S$ is under balancing selection is $T_2 = 2\{\ln[\mathcal{L}_{\text{M}}(\mathbf{n}, \boldsymbol{\rho}, \widehat{x} \,;\, \mathbf{a})] - \ln[\mathcal{L}_{\text{B}}(\mathbf{n} \,;\, \mathbf{a})]\}$. The two new methods, $T_1$ and $T_2$, have been implemented in the software package \textsc{ballet} (BALancing selection LikElihood Test), which is written in C.

            \subsection*{Evaluating the methods using simulations}
                \textcolor{black}{To evaluate the performance of $T_1$ and $T_2$ relative to HKA and Tajima's $D$, we carried out extensive simulations of balancing selection using different selection and demographic parameters. We simulated genomic data for a pair of species that diverged $\tau_D$ years ago. We introduced a site that is under balancing selection at time $\tau_S$, and the mode of balancing selection at the site is overdominance with selection strength $s$ and dominance parameter $h$. In the simulations discussed in this article, we varied the demographic history in the target ingroup species, the strength of selection $s$, the dominance parameter $h$, and the time at which the selected allele arises $\tau_S$. Details of how the simulations were implemented are further described in the \textsl{Materials and Methods} section.}

                \subsubsection*{Selected allele arising in ingroup species}
\begin{figure*}[h]
            \centering
            \includegraphics[width=1.0\textwidth]{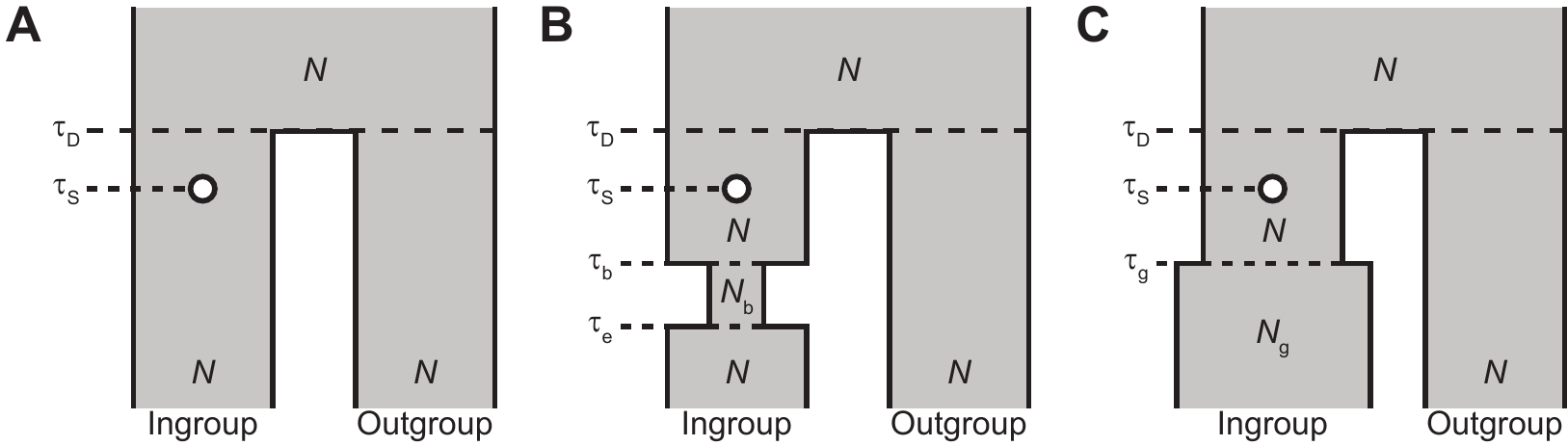}
            \caption{Demographic models used in simulations in which a selected allele arises after the split a pair of species. (\textsl{A}) Divergence model. Model parameters are a diploid effective population size $N$, divergence time $\tau_D$ of the ingroup and outgroup species, and the time $\tau_S$ when the selected allele arises. (\textsl{B}) Divergence model with a recent bottleneck within the ingroup species. Additional model parameters are the diploid effective population size $N_b$ during the bottleneck, the time $\tau_b$ when the bottleneck began, and the time $\tau_e$ when the bottleneck ended. (\textsl{C}) Divergence model with recent population growth within the ingroup species. Additional model parameters are the current diploid effective population size $N_g$ after recent growth and the time $\tau_g$ when the growth occurred.}
            \label{figure:simulation_models_ingroup_selection}
        \end{figure*}

\begin{figure*}[h]
            \centering
            \includegraphics[width=1.0\textwidth]{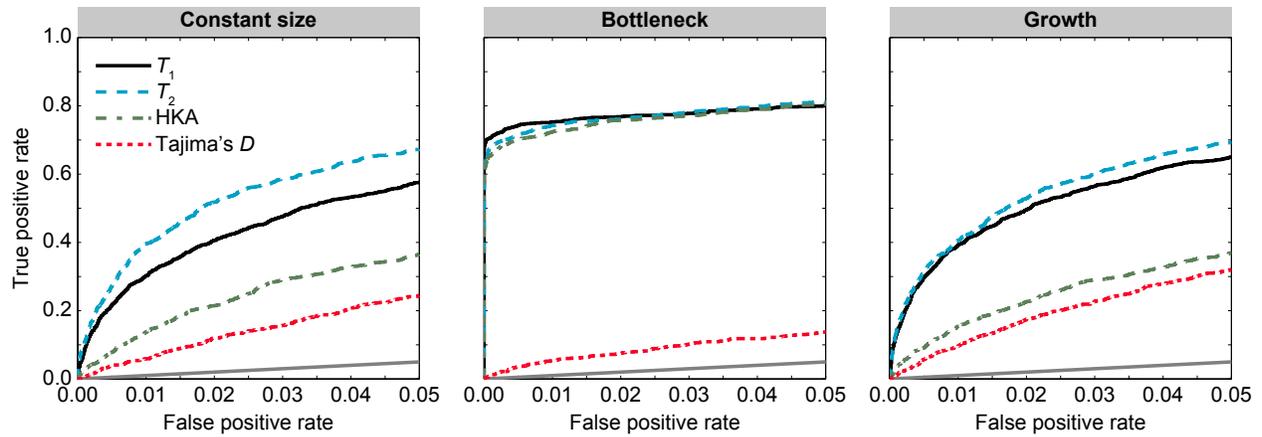}
            \caption{Performance of $T_1$, $T_2$, HKA, and Tajima's $D$ under the demographic models in Figure~\ref{figure:simulation_models_ingroup_selection} with selection parameter $s = 10^{-2}$ and dominance parameter $h = 100$. The first column is the divergence model in Figure~\ref{figure:simulation_models_ingroup_selection}\textsl{A}. The second column is the divergence model in Figure~\ref{figure:simulation_models_ingroup_selection}\textsl{B} with a recent bottleneck within the ingroup species. The third column is the divergence model in Figure~\ref{figure:simulation_models_ingroup_selection}\textsl{C} with recent population growth within the ingroup species.}
            \label{figure:simulation_results_s001_h100}
        \end{figure*}

                    We considered demographic models shown in Figure~\ref{figure:simulation_models_ingroup_selection} with $s = 10^{-2}$ and $h = 100$. For these simulations, we constructed receiver operator characteristic (ROC) curves, which illustrate the relationships between the true positive and false positive rates of the four methods. Figure~\ref{figure:simulation_results_s001_h100} displays ROC curves for $T_1$, $T_2$, HKA, and Tajima's $D$ under each of the three demographic models depicted in Figure~\ref{figure:simulation_models_ingroup_selection}, in which the strength of selection is $s = 10^{-2}$ and the dominance parameter is $h = 100$. This dominance parameter was chosen to represent an extremely strong level of heterozygote advantage. We later discuss a wider range of dominance parameters to test the limits of our methods. Under a model of constant population size (Fig.~\ref{figure:simulation_results_s001_h100}\textsl{A}), for a given false positive rate, $T_2$ tends to obtain more true positives than $T_1$, $T_1$ more true positives than HKA, and HKA more true positives than Tajima's $D$. In practice, however, we are typically concerned with a method's performance at low false positive rates. For a false positive rate of $1\%$, $T_1$, $T_2$, HKA, and Tajima's $D$ have true positive rates of 30, 40, 14, and $6\%$, respectively. Also, at a false positive rate of $5\%$, $T_1$, $T_2$, HKA, and Tajima's $D$ have true positive rates of 58, 67, 37, and $25\%$, respectively. These results show that $T_1$ and $T_2$ vastly outperform both HKA and Tajima's $D$, with $T_2$ performing better than $T_1$. However, the demographic model used in these simulations is the same as the one assumed in $T_1$ and $T_2$, namely, the standard neutral model. To examine the robustness of our methods, we considered two complex demographic scenarios that could potentially affect the results of our methods---a population bottleneck (Fig.~\ref{figure:simulation_models_ingroup_selection}\textsl{B}) and a population expansion (Fig.~\ref{figure:simulation_models_ingroup_selection}\textsl{C}).

                    Figure~\ref{figure:simulation_results_s001_h100}\textsl{B} displays ROC curves under a model in which the ingroup species experiences a recent severe bottleneck. Aside from Tajima's $D$, all of the methods perform well under this scenario. For a false positive rate of $1\%$, the true positive rates of $T_1$, $T_2$, HKA, and Tajima's $D$ are 75, 74, 72, and $5\%$, respectively. Similarly, for a false positive rate of $5\%$, the true positive rates of $T_1$, $T_2$, HKA, and Tajima's $D$ are 80, 81, 80, and $14\%$, respectively. This is because, under a model with a severe population bottleneck, there is a lower level of diversity across the genome and, hence, a lower polymorphism-to-substitution ratio. Because $T_1$, $T_2$, and HKA compare the level of polymorphism and divergence at a putatively selected site with that of the corresponding genome-wide background levels, these three methods identify a large excess of polymorphism compared to background levels at a site that is under balancing selection. However, Tajima's $D$ performs no such comparison and, thus, has little power to detect balancing selection under this scenario.

                    We next considered a demographic scenario in which the ingroup species experiences a recent population growth (Fig.~\ref{figure:simulation_models_ingroup_selection}\textsl{C}). Under this setting (Fig.~\ref{figure:simulation_results_s001_h100}\textsl{C}), similar to that of constant population size, $T_2$ tends to obtain more true positives than $T_1$, $T_1$ more true positives than HKA, and HKA more true positives than Tajima's $D$ for a given false positive rate. At a false positive rate of $1\%$, $T_1$, $T_2$, HKA, and Tajima's $D$ have true positive rates of 39, 41, 15, and $10\%$, respectively, and at a false positive rate of $5\%$, $T_1$, $T_2$, HKA, and Tajima's $D$ have true positive rates of 65, 69, 37, and $32\%$, respectively. Interestingly, all four methods perform better under a recent population growth than under a constant population size. This result is potentially due to more efficient selection after a population growth.
    				
                    By considering the demographic scenarios in Figure~\ref{figure:simulation_models_ingroup_selection}, we have demonstrated that our statistics, $T_1$ and $T_2$, generally outperform both HKA and Tajima's $D$. Additional simulation results are displayed in the \textsl{Supplementary Material}, in which we consider a range of values of the dominance parameter (\textsl{i.e.}, $h = 100$, 10, 3, and 1.5), a strong selection coefficient $s = 10^{-2}$ (Fig.~S1), a weak selection coefficient $s = 10^{-4}$ (Fig.~S2), and a scenario in which the selected allele arises in the population ancestral to the split of the ingroup and outgroup species (Fig.~S3-S5). In all scenarios tested, $T_1$ and $T_2$ perform as well as, though often better than, HKA and Tajima's $D$.

                    Next, we investigated scenarios in which we vary the dominance parameter $h$ with a selection coefficient of $s = 10^{-2}$. Considering an ingroup with a constant population size, $T_2$ outperforms $T_1$, $T_1$ outperforms HKA, and HKA outperforms Tajima's $D$ (Fig.~S1). As $h$ decreases, the performance of HKA and Tajima's $D$ decreases, yet the performance of $T_1$ and $T_2$ is not dramatically impacted. Hence, as $h$ decreases the performance of $T_1$ and $T_2$ relative to HKA and Tajima's $D$ increases, showing that the two new statistics provide a dramatic increase in power compared to HKA and Tajima's $D$.

                    Under a scenario in which the ingroup undergoes a recent population bottleneck, $T_1$, $T_2$, and HKA perform well, whereas Tajima's $D$ performs poorly (Fig.~S1). In addition, $h$ appears to have little influence on the relative performance of these methods. Hence, population bottlenecks tend to enhance the performance of $T_1$, $T_2$, and HKA, whereas they inhibit the performance of Tajima's $D$.

                    Moving to a scenario in which the ingroup undergoes a recent population expansion, Figure~S1 shows that $T_2$ outperforms $T_1$, $T_1$ outperforms HKA, and HKA outperforms Tajima's $D$.  The results in Figure~S1 indicate that the performance of $T_1$ is generally similar to $T_2$, whereas the performance of HKA and Tajima's $D$ is generally similar for large $h$ (\textsl{i.e}, $h = 10$ and 100), and dissimilar for low $h$ (\textsl{i.e}, $h = 1.5$ and 3). In addition, under the set of parameters investigated, $h$ appears to have little influence on the performance of $T_1$, $T_2$, and HKA, but causes the performance of Tajima's $D$ to decrease with decreasing $h$.

                    By considering a selection coefficient of modest strength (\textsl{i.e.}, $s = 10^{-2}$), we found that, in general, $T_1$ and $T_2$ perform quite well (Fig.~S1). However, as these two methods were developed to detect long-term balancing selection, then it is unclear how the methods should perform under a setting with weak selection. To investigate this scenario, we considered a weak selection coefficient of $s = 10^{-4}$, which is two orders of magnitude smaller than the one considered previously.

                    For a setting in which the ingroup remains at constant size, $T_2$ outperforms $T_1$, $T_1$ outperforms HKA, and HKA outperforms Tajima's $D$ (Fig.~S2) for large $h$ (\textsl{i.e.}, $h = 10$ and 100). In contrast to the results for the case of $s = 10^{-2}$, when $h$ is small (\textsl{i.e.}, $h = 1.5$ and 3), all methods perform poorly, each identifying signatures of selection only slightly better than random. Hence, when selection is weak and the level of overdominance is low, $T_1$ and $T_2$ cannot extract enough information from the data to create meaningful predictions. However, HKA and Tajima's $D$ perform just as poorly, and therefore $T_1$ and $T_2$ outperform HKA and Tajima's $D$ in general under a demographic model with constant population size.

                    Next, considering a situation in which the ingroup undergoes a recent population bottleneck, similarly to the observations for $s = 10^{-2}$, $T_1$, $T_2$, and HKA perform well, whereas Tajima's $D$ performs poorly (Fig.~S2). In contrast to the results for $s = 10^{-2}$, $h$ appears to have some influence on the relative performance of these methods. As $h$ decreases, the performance of all methods decreases---though not substantially. In addition, similarly to $s = 10^{-2}$, the performance of $T_1$, $T_2$, and HKA is approximately the same. Hence, even for weak selection, population bottlenecks tend to enhance the performance of $T_1$, $T_2$, and HKA, whereas they inhibit the performance of Tajima's $D$.

                    Finally, under a scenario in which the ingroup undergoes a recent population expansion, Figure~S2 shows that $T_2$ outperforms $T_1$, $T_1$ outperforms HKA, and HKA outperforms Tajima's $D$ for large $h$ (\textsl{i.e.}, $h = 10$ and 100).  In contrast to the results for the case of $s = 10^{-2}$, when $h$ is small (\textsl{i.e.}, $h = 1.5$ and 3), all methods perform poorly. Hence, like the case for an ingroup population with constant size, when selection is weak and the level of overdominance is low, $T_1$ and $T_2$ cannot extract enough information from the data to create meaningful predictions. However, HKA and Tajaima's $D$ perform just as poorly, and therefore $T_1$ and $T_2$ outperform HKA and Tajima's $D$ in general under a demographic model with recent population growth.

                \subsubsection*{Selected allele arising within ancestral population}
                    One hallmark of balancing selection is that it maintains polymorphism for a long time, potentially for millions of years \cite{KleinEtAl93, KleinEtAl98, KleinEtAl07}. Due to the extreme age of some balanced polymorphisms, they tend to occur within multiple species, thereby creating a polymorphism shared across species referred to as a trans-specific polymorphism. Figure~S3 displays the three models that we consider in which a selected allele arises in the population ancestral to the split of the ingroup and outgroup species. For each of the three demographic scenarios, we set $\tau_S = 1.5 \times 10^7$ years ago, creating a selected allele that is three times as ancient as the one that we consider in Figure~2. All other models parameters are identical to those considered in Figure~2.

                    Here we investigate the performance of $T_1$, $T_2$, HKA, and Tajima's $D$ in the context of demographic models in which a selected allele arises in an ancestral population and in which the selective pressure is of modest strength (\textsl{i.e.}, $s = 10^{-2}$) and varying dominance $h$. For a setting in which the ingroup remains at constant size, $T_2$ outperforms $T_1$, $T_1$ outperforms HKA, and HKA outperforms Tajima's $D$ (Fig.~S4). As $h$ decreases, the performance of HKA and Tajima's $D$ decreases, yet the performance of $T_1$ and $T_2$ is not dramatically impacted. Hence as $h$ decreases the performance of $T_1$ and $T_2$ relative to HKA and Tajima's $D$ increases, mirroring the results observed in Figure~S1.

                    Next, considering a situation in which the ingroup undergoes a recent population bottleneck, $T_1$, $T_2$, and HKA perform well, whereas Tajima's $D$ performs poorly (Fig.~S4). In addition, $h$ appears to have little influence on the relative performance among $T_1$, $T_2$, and HKA yet causes Tajima's $D$ to perform worse for small $h$ (\textsl{i.e.}, $h = 1.5$). Hence, akin to the observations for Figure~S1, population bottlenecks tend to enhance the performance of $T_1$, $T_2$, and HKA, whereas they inhibit the performance of Tajima's $D$.

                    Under a scenario in which the ingroup undergoes a recent population expansion, Figure~S4 shows that in most cases $T_2$ outperforms $T_1$, $T_1$ outperforms HKA, and HKA outperforms Tajima's $D$. The performance of $T_1$ is generally similar to $T_2$, whereas the performance of HKA and Tajima's $D$ generally similar for large $h$ (\textsl{i.e}, $h = 10$ and 100), and dissimilar for low $h$ (\textsl{i.e}, $h = 1.5$ and 3). Interestingly, for $h = 1.5$, $T_1$ performs slightly better than $T_2$. In addition, under the set of parameters investigated, $h$ appears to have little influence on the performance of $T_1$, $T_2$, and HKA, but causes the performance of Tajima's $D$ to decrease with decreasing $h$.

                    By considering the demographic model in Figure~S3, we have shown that the performance of $T_1$, $T_2$, HKA, and Tajima's $D$ are not greatly impacted by the age of the selected allele, provided that the selected allele is old and has maintained balancing selection for an extended period of time. Hence, though $T_1$ and $T_2$ make the assumption that lineages from the ingroup species are monophyletic, this assumption does not hinder the methods in practice.

                    For a setting in which the ingroup remains at constant size, $T_2$ outperforms $T_1$, $T_1$ outperforms HKA, and HKA outperforms Tajima's $D$ (Fig.~S5) for large $h$ (\textsl{i.e.}, $h = 10$ and 100). In contrast, when $h$ is small (\textsl{i.e.}, $h = 1.5$ and 3), all methods perform poorly, each identifying signatures of selection only slightly better than random. Hence, as observed in Figure~S2, when selection is weak and the level of overdominance is low, $T_1$ and $T_2$ cannot extract enough information from the data to create meaningful predictions.

                    Next, considering a situation in which the ingroup undergoes a recent population bottleneck, $T_1$, $T_2$, and HKA perform well, whereas Tajima's $D$ performs poorly (Fig.~S5). In addition, $h$ appears to have some influence on the relative performance of these methods. As $h$ decreases, the performance of all methods decreases---though not substantially. Also, the performance of $T_1$, $T_2$, and HKA is approximately the same. These results mirror those observed in Figure~S2, and thus, even for weak selection at a trans-specific polymorphism, population bottlenecks tend to enhance the performance of $T_1$, $T_2$, and HKA, whereas they inhibit the performance of Tajima's $D$.

                    Finally, under a scenario in which the ingroup undergoes a recent population expansion, Figure~S5 shows that $T_2$ outperforms $T_1$, $T_1$ outperforms HKA, and HKA outperforms Tajima's $D$ for large $h$ (\textsl{i.e.}, $h = 10$ and 100).  In contrast, when $h$ is small (\textsl{i.e.}, $h = 1.5$ and 3), all methods perform poorly. Hence, like the case for an ingroup population with constant size, when selection is weak and the level of overdominance is low, $T_1$ and $T_2$ cannot extract enough information from the data to create meaningful predictions.

                    These results show that, for the case of weak selection, a setting in which the selected allele generates trans-specific polymorphisms has little effect on the performance on $T_1$, $T_2$, HKA, and Tajima's $D$ when compared with their respective performances under the case in which the polymorphism is not trans-specific. Hence, we have shown that the performance of $T_1$ and $T_2$ is not influenced by the presence of a trans-specific polymorphism even though they are based on the assumption that lineages from the ingroup species are monophyletic.

            \subsection*{Empirical analysis}
                \subsubsection*{Balancing selection in humans}
                    We probed the effects of balancing selection in humans by using whole-genome sequencing data from nine unrelated individuals from the CEU population and nine unrelated individuals from the YRI population (see \textsl{Materials and Methods}). We performed a scan for balancing selection at each position in our dataset by considering a window of 100 substitutions or polymorphisms upstream and downstream of our focal site. This window size was taken for computational convenience, rather than by consideration of the recombination rate or polymorphism density within the region. Though we used a window size of 200 polymorphisms or substitutions for computational convenience, $T_1$ and $T_2$ can also be computed using all sites on a chromosome. The mean window length was $\sim$14.7kb for the CEU and $\sim$13.7kb for the YRI populations, which should be sufficiently long because recombination quickly breaks down the signal of balancing selection at distant neutral sites. Manhattan plots for $T_1$ (Figs.~S6~and~S7) and $T_2$ (Figs.~S8~and~S9) test statistics suggest that there are multiple outlier candidate regions. Intersecting the locations of these scores with those from the longest transcript of each RefSeq gene (\textsl{i.e.}, coding region) led to identification of many previously-hypothesized and novel genes potentially undergoing balancing selection (see Tables~S1-S4, with previously-hypothesized genes highlighted in bold).

\begin{figure*}[h]
            \centering
            \includegraphics[width=1.0\textwidth]{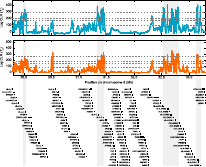}
            \caption{Signals of balancing selection within the HLA region for the CEU (blue) and YRI (orange) populations using the $T_2$ test statistic. From bottom to top, the horizontal dotted gray lines indicate the $0.5\%$, $0.1\%$, $0.05\%$, and $0.01\%$ empirical cutoffs, respectively.}
            \label{figure:hla_spect}
        \end{figure*}

                    Multiple genes at the HLA region are strong outliers (top $0.01\%$ of all scores across the genome) in our scan for balancing selection (Tables~S1-S4). Because this study uses high-coverage sequencing data, resolution in the HLA region is particularly fine (Figs.~S10~and~\ref{figure:hla_spect}), with  strong signals in classical MHC genes such as \textsl{HLA-A}, \textsl{HLA-B}, \textsl{HLA-C}, \textsl{HLA-DR}, \textsl{HLA-DQ}, and \textsl{HLA-DP} genes \cite{BubbEtAl06}. The HLA region, which is located on chromosome six, is a well-known site of balancing selection in humans \cite{KleinEtAl93, KleinEtAl98, KleinEtAl07}. The protein products encoded by HLA genes are involved in antigen presentation, thus playing important roles in immune system function. Genes at the HLA locus are known to be highly polymorphic and are thought to be subject to balancing selection due to frequency-dependent selection, overdominance, or fluctuating selection in a rapidly changing pathogenic environment \cite{TakahataAndNei90, Hedrick02}. As the HLA region is so well known as a locus under balancing selection, it is important that our methods identify strong candidate candidate genes in the regions as a proof of concept.

\begin{figure*}[h]
            \centering
            \includegraphics[width=1.0\textwidth]{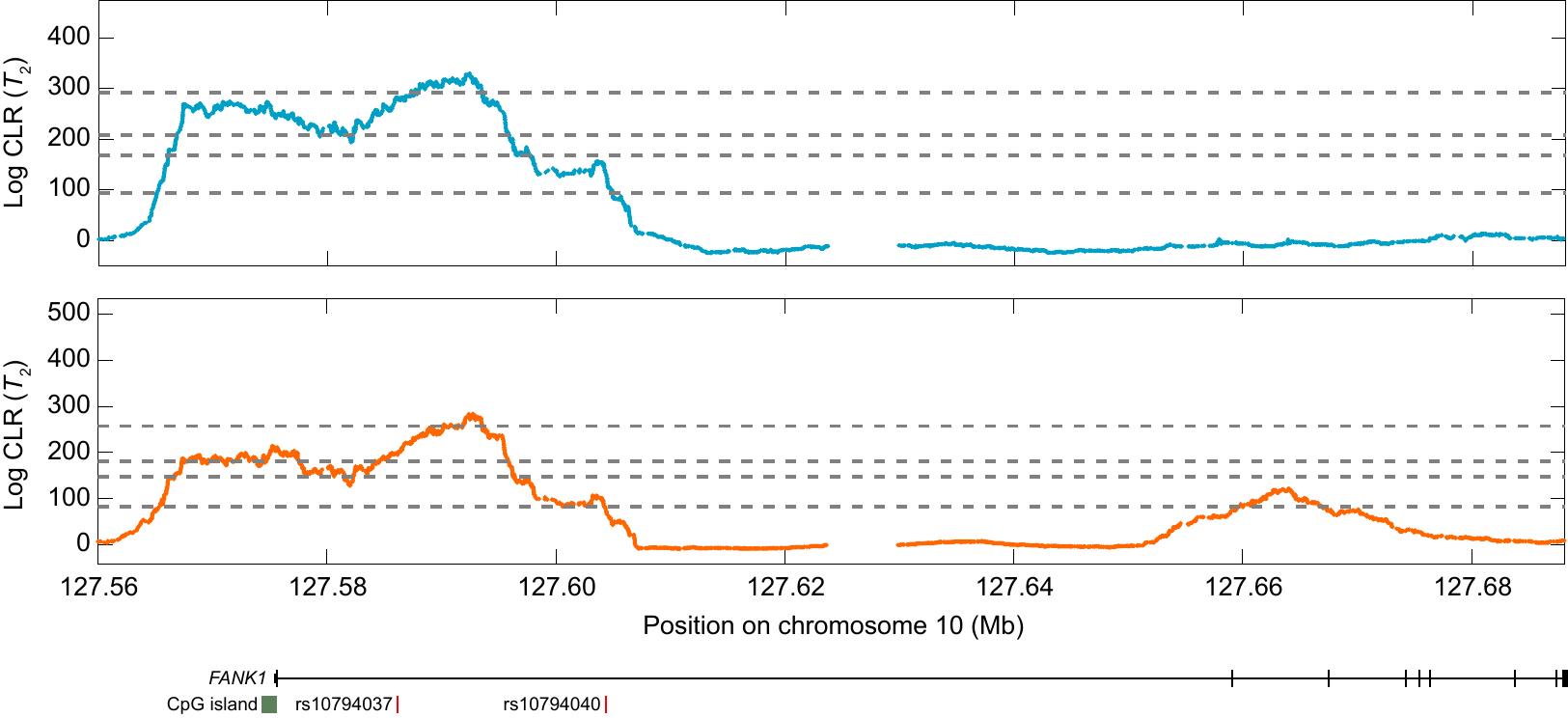}
            \caption{Signal of balancing selection at the \textsl{FANK1} gene for the CEU (blue) and YRI (orange) populations using the $T_2$ test statistic. From bottom to top, the horizontal dotted gray lines indicate the $0.5\%$, $0.1\%$, $0.05\%$, and $0.01\%$ empirical cutoffs, respectively. SNPs (rsIDs) correspond to markers showing significant levels of transmission distortion within the Meyer~\textsl{et al.} study \cite{MeyerEtAl12}.}
            \label{figure:fank1_spect}
        \end{figure*}

                    One gene that we found particularly intriguing is \textsl{FANK1} (Figs.~S11~and~\ref{figure:fank1_spect}). This gene is one of the top four candidates in the CEU and YRI populations when using either the $T_1$ or $T_2$ statistic (Tables~S1-S4). In addition, \textsl{FANK1} is the top candidate among genes that have not been previously hypothesized to be under balancing selection when using either test in the CEU and the $T_1$ test in the YRI. \textsl{FANK1} is expressed during the transition from diploid to haploid state in meiosis \cite{ZhengEtAl07, WangEtAl11}. Though it is often identified as spermatogenesis-specific \cite{ZhengEtAl07, WangEtAl11}, it is also expressed during oogenesis in cattle \cite{HwangEtAl05} and mice \cite{ZuccottiEtAl08}. Its function is to suppress apoptosis \cite{WangEtAl11}, and it is one of ten to 20 genes identified as being imprinted in humans (\textsl{i.e.}, allele specific methylation) \cite{LiEtAl10}. Interestingly, it also shows marginal evidence of segregation distortion (Fig.~\ref{figure:fank1_spect}) \cite{MeyerEtAl12}. Further, as a CpG island resides directly underneath our signal in both the CEU and YRI populations, we analyzed the region around \textsl{FANK1} with all $GC \rightarrow AT$ transitions on chromosome 10 removed as well as all transitions on chromosome 10 removed and we still retain the peak (Fig.~S12), strongly suggesting that the signature of balancing selection that we identified around \textsl{FANK1} is not driven by CpG mutational effects.

                \subsubsection*{Gene ontology analysis}
                    To elucidate functional similarities among genes identified to be under balancing selection, we performed gene ontology (GO) enrichment analysis using \textsc{GOrilla}\cite{EdenEtAl07, EdenEtAl09}. First, we compared an unranked list of the top 100 candidate genes (Tables~S1-S4) to the background list of all unique genes. Genes obtained using either test statistic are enriched for processes involved in the immune response in both the CEU and YRI populations (Tables~S5-S8). Similarly, the top genes are enriched for MHC class II functional categories (Tables~S9-S12), with the exception of the $T_2$ statistic applied to YRI, which has no functional enrichment (Table~S12). Further, these top genes tend to be components of the MHC complex and membranes (Tables~S13-S16), which often directly interact with pathogens. Interestingly, removing all HLA genes from both the top 100 and background sets of genes reveals no GO enrichment for process, function, or component categories, indicating that enrichment is predominately driven by the HLA region. Because we can also provide a score for each candidate gene in our likelihood framework, we performed a second analysis in which we ranked genes by their likelihood ratio test statistic, with the goal of identifying GO categories that are enriched in top-ranked genes. Using this framework, the top candidate genes tend to be involved in immune response and cell adhesion processes (Tables~S17-S20); MHC activity and membrane protein activity functions, such as transporting and binding molecules (Tables~S21-S24); and MHC complex, membrane, and cell junction components (Tables~S25-S28).  In contrast to the case of the top 100 candidate genes, removing all HLA genes from the ranked list still resulted in GO enrichment in categories such as cell adhesion (processes), membrane protein activity (function), and components of membranes and cell junctions (component).

        \section*{Discussion}
            In this article, we presented two likelihood-based methods,  $T_1$ and $T_2$, to identify genomic sites under balancing selection. These methods combine intra-species polymorphism and inter-species divergence with the spatial distribution of polymorphisms and substitutions around a selected site. Through simulations, we showed that $T_1$and $T_2$ vastly outperform both the HKA test and Tajima's $D$ under a diverse set of demographic assumptions, such as a population bottleneck and growth. In addition, application of  $T_1$ and $T_2$ to whole-genome sequencing data from Europeans and Africans revealed many previously identified and novel loci displaying signatures of balancing selection.

            Simulation results suggest that $T_2$ performs at least as well as $T_1$, and so a natural question is whether $T_1$ would ever be used. Based on the fact that $T_2$ uses the allele frequency spectrum and $T_1$ does not, then $T_1$ would be a valuable statistic to employ when allele frequencies cannot be estimated well. \textcolor{black}{One example is a situation in which the sample size is small (e.g., one or two genomes). Under this scenario, the $T_2$ test statistic would likely provide little additional power over the $T_1$ statistic.} As another example, it is becoming increasingly common for studies to sequence a pooled sample of individuals rather than each individual in the sample separately. This pooled sequencing will tend to yield inaccurate estimates of allele frequencies across the genome, which could heavily influence the performance of the $T_2$ statistic. However, if there is sufficient enough evidence that a site has a pair of alleles observed in the sample, then this site can be considered polymorphic regardless of its actual allele frequency. Future developments that can statistically account for this uncertainty in allele frequency estimation could be incorporated into the $T_2$ test statistic so that it can be applied to pooled sequencing data.

            The model of balancing selection used in this article is from Hudson~and~Kaplan \cite{HudsonAndKaplan88}, and assumes that natural selection is so strong that it maintains a constant allele frequency at the selected locus forever. The simulation scenarios considered here assumed that the strength of balancing selection was also constant since the selected allele arose. However, selection coefficients can fluctuate over time, which provides the basis for future work on investigating the robustness of methods for detecting balancing selection under scenarios in which the strength of selection fluctuates or when selection is weak. Future work can use the framework developed here to construct methods for identifying balancing selection under models with more relaxed assumptions (e.g., see Barton~and~Etheridge \cite{BartonAndEtheridge04} and Barton~\textsl{et al}. \cite{BartonEtAl04} for potential models).

            Though we have shown that $T_1$ and $T_2$ perform well under a population bottleneck and growth, they may be less robust to other forms of demographic model violations, such as population structure. Because population subdivision increases the time to coalescence and corresponding length of a genealogy, we expect higher levels of polymorphism across the genome. Under most assumptions, population subdivision affects the genome uniformly;  it increases the level of background polymorphism and likely only slightly decreases the power of the new statistics. However, in some cases, such as an ancient admixture event (e.g., with Neanderthals\cite{GreenEtAl10} or Denisovans\cite{ReichEtAl10}), levels of variability may increase in only a few regions of the genome, increasing the mean coalescence time in these regions. Such regions may appear to have excess polymorphism relative to background levels and, hence, display false signals of balancing selection under the $T_1$ statistic.  However, in non-African humans, introgressed regions typically have low population frequencies\cite{GreenEtAl10, ReichEtAl10}, and, hence, it would be unlikely for polymorphic sites in these regions to harbor many introgressed alleles segregating at intermediate frequencies. Thus, the $T_2$ statistic, which explicitly utilizes allele frequency spectra information, would likely be able to distinguish these blocks of archaic admixture from regions of balancing selection. Further, as observed in other studies of natural selection \cite{JensenEtAl07, PavlidisEtAl10}, increased robustness to confounding demographic processes can potentially be gained through the use of additional information. For example, population bottlenecks as well as gene flow can increase linkage disequilibrium \cite{PlagnolAndWall06, Slatkin08}. Therefore, knowledge about linkage disequilibrium in a region could aid in distinguishing population subdivision from long-term balancing selection.

            Another concern when performing genomes scans for balancing selection is the possibility of false positives due to bioinformatical errors.  For example, misalignment of sequence reads in duplicated regions may lead to falsely elevated levels of variability.  In many cases, this problem can be alleviated by removing duplicated regions from analyses. However, a non-negligible portion of the human genome is not represented in standard reference sequences and, thus, there may be many unidentified paralogs in the genome.  Fortunately, removing sites that deviate from Hardy-Weinberg equilibrium  helps to alleviate these problems, because SNPs fixed between or segregating at high frequencies in one of two (or more) paralogous regions will have an excess of heterozygotes in  combined short-read alignments. We applied a Hardy-Weinberg filter to all empirical data analyzed in this article. We note that deviations from Hardy-Weinberg equilibrium are expected under certain forms of balancing selection. In theory, a balancing selection signal could, therefore, be lost due to such filtering. However, we used a filtering cutoff of $p < 10^{-4}$ (see \textsl{Materials and Methods}). The strength of selection required to cause this type of deviation from Hardy-Weinberg equilibrium used in the filtering is extremely strong, and such selection would almost certainly have been detected using other methods. Well-established examples of balancing selection in the human genome, such as the selection affecting the HLA loci, are not lost because of filtering, and would generally not be easily detectable using deviations from Hardy-Weinberg as a test. Nonetheless, because phenomena other than balancing selection, such as bioinformatical errors or archaic admixture, could potentially lead to false signals of balancing selection, additional evidence should be obtained before definitively concluding that a site has been subjected to balancing selection.

            One source of additional evidence of balancing selection is whether a signal lies within a region harboring a trans-specific polymorphism \cite{SegurelEtAl12, LefflerEtAl13} because it is unlikely to have a polymorphism segregating in each of a pair of closely-related species without selection maintaining the polymorphism. However, relying solely on evidence from trans-specific polymorphisms would miss many true signals of balancing selection that are not maintained as trans-specific polymorphisms. In addition, regions with bioinformatical errors (e.g., mapping errors) may give the same errors in both species, resulting in a false signal of a shared polymorphism between the pair of species. Nevertheless, the observation of a trans-specific polymorphism can provide convincing evidence of ancient balancing selection \cite{SegurelEtAl12, LefflerEtAl13}. Previous studies of selection have shown that combinations of statistics can be powerful tools when identifying genes under selection \cite{Innan06, AndresEtAl09, GrossmanEtAl10}. Hence, combining our methods with other summaries (e.g., linkage disequlibrium \cite{PlagnolAndWall06, JensenEtAl07, Slatkin08, PavlidisEtAl10}) or information on trans-species polymorphisms \cite{SegurelEtAl12, LefflerEtAl13} will lead to increasingly effective approaches for detecting balancing selection.

            Another commonly-cited source of evidence for balancing selection is based on consideration of the topology and branch lengths of within-species haplotype trees. Under long-term balancing selection, the underlying genealogy (e.g., see Fig.~S13) will be symmetric, with long basal branches separating a pair of allelic classes (\textsl{i.e.}, haplotypes containing one variant and haplotypes containing the other variant). However, the underlying genealogy for a linked neutral variant may differ substantially from that of the selected site. Around a balanced polymorphism, there will be a strong reduction of linkage disequilibrium, not unlike a recombination hotspot, because the long genealogy in the balanced polymorphism provides extra opportunities for recombination. Consequently, the signal of balancing selection will be narrow, and trees estimated from sites located in a  window  around the balanced polymorphism may fail to detect the presence of highly divergent haplotypes. The utility of within-species haplotype trees as a signature of long-term balancing selection is unclear, as the genealogy of the haplotype may not match the genealogy of the selected region.  For example, Figure~S14 shows that haplotype trees based on scenarios under balancing selection appear similar to those under neutrality, with the difference that external branches are slightly longer under balancing selection than under neutrality, which contrasts with the generally-held belief that basal branches should be long. As such, haplotype networks or trees may not be powerful tools for identifying regions under balancing selection.

            Within our scan, we identified a gene called \textsl{FANK1}, which is expressed during the transition from diploid to haploid states in meiosis \cite{ZhengEtAl07, WangEtAl11}, is often identified as spermatogenesis-specific \cite{ZhengEtAl07, WangEtAl11}, suppresses apoptosis \cite{WangEtAl11}, is imprinted \cite{LiEtAl10}, and exhibits evidence of segregation distortion (Fig.~\ref{figure:fank1_spect}) \cite{MeyerEtAl12}. These characteristics suggest that maintenance of polymorphism at \textsl{FANK1} results from segregation distortion, which can occur when the allele favored by distortion is associated with negative fitness effects, particularly if the negative effect is pronounced in the homozygous state (see p.~562-563 of Charlesworth~and~Charlesworth \cite{CharlesworthAndCharlesworth10}; \'{U}beda~and~Haig \cite{UbedaAndHaig04}). The distorting allele will increase in frequency when rare because of the segregation distortion in heterozygotes.  But when it becomes common, selection will act against it because it will more often occur in the homozygous state when rare.  Under such a scenario, theoretical results suggest that it is possible for a distorter to spread through a population without reaching fixation, obtaining a frequency that permits the maintenance of a stable polymorphism (see p.~564 of Charlesworth~and~Charlesworth \cite{CharlesworthAndCharlesworth10}). In addition, the inclusion of imprinting at such a locus further enchances the parameter space at which a polymorphism can be maintained \cite{UbedaAndHaig04}.

            The function of \textsl{FANK1} makes it a particularly good candidate for harboring alleles causing segregation distortion.  It is expressed primarily during meiosis and inhibits apoptosis, which has previously been hypothesized to be associated with segregation distortion \cite{NielsenEtAl05PLoSBiology, daFonsecaEtAl10}.  A large proportion of sperm cells are eliminated by apoptosis, so allelic variants causing avoidance of apoptosis after meiosis could serve as segregation distorters. However, mutations that lead to avoidance of apoptosis may be associated with negative fitness effects, especially in the homozygous states, because  they could lead to dysspermia or azoospermia.  Apoptosis during spermatogenesis plays a critical role in maintaining the optimal relationship between the number of developing sperm cells and sertoli cells, which support developing sperm cells.

            Though some of the sites identified in \textsl{FANK1} show marginal levels of segregation distortion, the region displaying the largest level of segregation distortion in the human genome is located 300kb upstream of \textsl{FANK1}\cite{MeyerEtAl12}. Further, a recent genome-wide association study for male fertility identified a significant SNP (rs9422913) located approximately 250kb upstream of \textsl{FANK1}\cite{KosovaEtAl12}. Even though these regions are quite distant from \textsl{FANK1}, if strong enough linkage exists with \textsl{FANK1}, then it is possible for a two-locus segregation distorter to spread within a population (p.~569 of Charlesworth~and~Charlesworth \cite{CharlesworthAndCharlesworth10}). Hence the signals of segregation distortion\cite{MeyerEtAl12} and fertility\cite{KosovaEtAl12} displayed in these regions upstream of \textsl{FANK1} could be a result of an association with \textsl{FANK1}.

            Thus, \textsl{FANK1} is an interesting candidate for further study of balancing selection. The association of segregation distortion and balancing selection has been empirically described in other species, e.g., \textsl{Caenorhabditis elegans}\cite{SeidelEtAl08}. However, as it has not yet been documented in humans, \textsl{FANK1} may be the first example of a segregation distorter causing balancing selection in humans.

            In the last several years, there has been an accumulation of evidence against the pervasiveness of hard sweeps in some species, e.g., in humans \cite{HernandezEtAl11, LohmuellerEtAl11PLoS, GrankaEtAl12}. Instead, other adaptive forces, such as balancing selection, could play an important role in shaping genetic variation across the genome. Interestingly, a recent theoretical study showed that a large proportion of adaptive mutations in diploids leads to heterozygote advantage\cite{SellisEtAl12}, suggesting that much of the genome may be under balancing selection. If this intriguing prospect is true, then because our methods for detecting balancing selection are the most powerful that have been developed to date, they will be useful tools in uncovering the potentially many regions under balancing selection in humans and other species.

        \section*{Materials and Methods}
            \subsection*{A simple estimate of the inter-species expected coalescence time}
                \textcolor{black}{For the purposes of our simulation and empirical analyses, we introduce a basic estimate of the expected coalescence time between the ingroup and outgroup species. Consider a sample of $n$ lineages (\textsl{i.e.}, $n$ haploid individuals) from an ingroup species and one lineage from an outgroup species. For simplicity, assume that the ingroup species, outgroup species, and ancestral species from which the ingroup and outgroup diverged has an effective population size of $N = 10^4$ diploid individuals. Further, assume that the per-site per-generation mutation rate is $\mu = 2.5 \times 10^{-8}$ and that the total sequence length analyzed is $K$. We estimate the expected coalescence time of all $n$ lineages in the ingroup species as $\widehat{H} = \widehat{\pi} / [4N\mu K (1 - 1/n)]$, where $\widehat{\pi}$ is the mean number of pairwise sequence differences and $4N\mu K (1 - 1/n)$ is the expected number of mutations for a sequence of length $K$ and $n$ sampled lineages. Suppose that $\widehat{d}$ is the number of substitutions of fixed differences observed between the intgroup and outgroup species. Then we estimate the mean coalescence time between the ingroup and outgroup species by $\widehat{C} = [\widehat{H} + \widehat{d}/(2N\mu K)]/2$.}

            \subsection*{Application of the new test statistics to data}
                In the empirical analysis of human genomic data, we obtained values for the $T_1$ and $T_2$ test statistics for a large number of positions spaced across the genome. From these values, we overlapped protein coding region (or genes) with the positions in the genome that the test statistics were calculated at. We assigned the value of the test statistic for the gene as the maximal value of the test statistic for the positions that it overlapped. We then ranked the set of genes based on their scores to identify genes that are outliers. Note that we are not attempting to identify regions with statistical significance or a certain $p$-value threshold, but instead are looking for genes that may be outliers, and so the $0.01$, $0.05$, $0.10$, and $0.50\%$ empirical cutoffs are not meant to represent a formal significance cutoff.

                When applying the $T_1$ and $T_2$ test statistics to simulated and empirical data, we do not estimate the rate of mutation $\theta_1$ from $A_1$ alleles to $A_2$ alleles or the rate of mutation $\theta_2$ from $A_2$ alleles to $A_1$ alleles at the selected site $S$, as defined in the Hudson-Darden-Kaplan model. We instead treat these rates as a constant, with $\theta_1 = \theta_2 = 0.05$ for the analyses in this article. The motivation is that, if these mutation rates did not exist, then the tree height would increase rapidly for small recombination rates. Our method assumes that a most recent common ancestor of the set of sampled alleles is reached more recently than the inter-species coalescence time $\widehat{C}$ between the ingroup and outgroup species (\textsl{i.e.}, $H_n(x, \rho) < \widehat{C}$ even for small $\rho$). Simulation results (see \textsl{Evaluating the methods using simulations}) show that our new methods perform extremely well, even though we set the nuisance $\theta_1$ and $\theta_2$ parameters to a constant value. To maximize of the equilibrium frequency $x$ of the $A_1$ allele, we utilized the value of $x$, denoted by $\widehat{x}$, that maximized the composite likelihood under the model, by choosing $\widehat{x}$ from values of $0.05, 0.10, \ldots, 0.95$.

            \subsection*{Simulation procedure to evaluate the performance of $T_1$ and $T_2$}
                We applied $T_1$ and $T_2$ to data simulated under population divergence models, using parameters to mimic humans (ingroup) and chimpanzees (outgroup). The models that we simulated under are illustrated in Figure~\ref{figure:simulation_models_ingroup_selection}. For each of three models, we set each of the ingroup, outgroup, and ancestral population sizes to $N = 10^4$ diploid individuals \cite{TakahataEtAl95} and the divergence time between the ingroup and the outgroup species to $\tau_D = 5 \times 10^6$ years ago \cite{KumarEtAl05}. We assumed a generation time of 20 years \cite{NachmanAndCrowell00}, a mutation rate of $\mu = 2.5 \times 10^{-8}$ mutations per-nucleotide per-generation \cite{NachmanAndCrowell00}, a recombination rate of $r = 2.5 \times 10^{-8}$ recombinations per-nucleotide per-generation, and a sequence length of $10^5$ nucleotides. Assuming a per-generation selection coefficient $s$, where $0 \leq s \leq 1$,\ and a dominance parameter $h$, where $h > 1$, at time $\tau_S$, a selected allele arose and evolved under an overdominance model with $A_1A_1$ homozygotes having fitness 1, $A_1A_2$ heterozygotes having fitness $1 + hs$, and $A_2A_2$ homozygotes having fitness $1 + s$. Simulations were performed using \textsc{mpop} \cite{PickrellEtAl09}, which was seeded with population-level chromosome data generated by the neutral coalescent simulator \textsc{ms} \cite{Hudson02}. After the completion of each simulation, we sampled 18 chromosomes from the ingroup species and one chromosome from the outgroup species. Ancestral alleles were called using the outgroup species, and so the called ancestral allele may not actually be the true ancestral allele. For each of the three demographic scenarios, we set $\tau_S = \tau_D = 5 \times 10^6$ years ago. For the bottleneck model (Fig.~\ref{figure:simulation_models_ingroup_selection}\textsl{B}), we set the bottleneck population size to $N_b = 550$ diploid individuals, the time at which the bottleneck began to $\tau_b = 3.0 \times 10^4$ years ago, and the time at which the bottleneck ended to $\tau_e = 2.2 \times 10^4$ years ago \cite{LohmuellerEtAl09, LohmuellerEtAl11}. For the growth model (Fig.~\ref{figure:simulation_models_ingroup_selection}\textsl{C}), we set the expanded population size to $N_g = 2 \times 10^4$ diploid individuals and the time at which the population began to grow to $\tau_g = 4.8 \times 10^4$ years ago \cite{LohmuellerEtAl11}.

            \subsection*{Empirical dataset construction}
                We used data from nine European and nine African diploid genomes sequenced by Complete Genomics \cite{DrmanacEtAl09}. All individuals were unrelated \cite{PembertonEtAl10}, with the European individuals from the CEU population (NA06985, NA06994, NA07357, NA10851, NA12004, NA12889, NA12890, NA12891, NA12892) and the African individuals from the YRI population (NA18501, NA18502, NA18504, NA18505, NA18508, NA18517, NA19129, NA19238, NA19329). We used the genotype calls made by Complete Genomics that were found in the ``masterVarBeta'' files. We downloaded pairwise alignments between human reference hg18 and chimpanzee reference panTro2 from the UCSC Genome Browser at http://genome.ucsc.edu/. Sites with more than two distinct alleles across all Complete Genomics individuals as well as the hg18-panTro2 alignments, sites in the Complete Genomics data where one of the two alleles did not match the reference sequence, and sites that were within two nucleotides of structural variants called in any one of the Complete Genomics individuals were removed. In addition, combining all 54 unrelated individuals in the Complete Genomics dataset, sites that had a $p$-value less than $10^{-4}$ for a one-tailed Hardy-Weinberg test of excess heterozygotes \cite{WiggintonEtAl05} were excluded. It should be noted that under a scenario of heterozygote advantage, it is expected that we should observe an excess of heterozygous individuals at sites in the vicinity of the site under balancing selection. However, a major concern with sequencing data are mapping errors, and so the Hardy-Weinberg filter is necessary to reduce the confounding effects of regions with these bioinformatical artifacts. As a consequence, this filter may increase the chance that we miss certain regions that are under balancing selection in our scan. Finally, sites that were polymorphic in the Complete Genomics sample (\textsl{i.e.}, either CEU or YRI) and sites that contained a fixed difference between the Complete Genomics sample and the chimpanzee reference sequence were retained. As in the simulations, the ancestral allele was called using the chimpanzee outgroup, and so the called ancestral allele may not be the true ancestral allele. However, simulation results shows that our new methods perform well even when the ancestral allele is potentially misspecified. Further, it may be possible to account for ancestral allele misspecification by using multiple outgroups, or by statistically accounting for the misspecification \cite{HernandezEtAl07}.

                To obtain recombination rates between pairs of sites, we used the sex-averaged pedigree-based human recombination map from deCODE Genetics \cite{KongEtAl10}. We constructed recombination rates between all pairs of sites in the filtered Complete Genomics samples by linearly interpolating rates between adjacent sites within the sex-averaged maps.

        \section*{Acknowledgments}
            We thank Zachary Szpiech for coming up with the name \textsc{ballet} and Zelia Ferreira for help testing early versions of \textsc{ballet}. This material was supported by National Science Foundation grant DBI-1103639 (MD), a Miller Research Fellowship from the Miller Research Institute at the University of California, Berkeley (KEL), and National Institutes of Health grant 3R01HG03229-07 (RN).

        \bibliography{References}

        \clearpage

        \clearpage

        \clearpage

        \clearpage

        \clearpage

    \end{sloppypar}

    \bibliographystyle{plos2009}
\end{document}